\documentclass[12pt,preprint]{aastex}
\def\hst{{\it Hubble Space Telescope}}

\shorttitle{Morphologies of GRB Host Galaxies}
\shortauthors{Wainwright et al.}

\def\nod{\nodata}

\def\pom{1}
\def\ociw{2}
\def\prince{3}
\def\hubble{4}

\begin{document}

\title{\large A Morphological Study of Gamma-Ray Burst Host Galaxies}

\author{
C.~Wainwright\altaffilmark{\pom},
E.~Berger\altaffilmark{\ociw,}\altaffilmark{\prince,}\altaffilmark{\hubble},
B.~E.~Penprase\altaffilmark{\pom},
}

\altaffiltext{\pom}{Pomona College Department of Physics and Astronomy, 
610 N. College Avenue, Claremont, CA}

\altaffiltext{\ociw}{Observatories of the Carnegie Institution
of Washington, 813 Santa Barbara Street, Pasadena, CA 91101}
 
\altaffiltext{\prince}{Princeton University Observatory,
Peyton Hall, Ivy Lane, Princeton, NJ 08544}
 
\altaffiltext{\hubble}{Hubble Fellow}

\begin{abstract} 

We present a comprehensive study of the morphological properties of 42
$\gamma$-ray burst (GRB) host galaxies imaged with the \hst\ in the
optical band.  The purpose of this study is to understand the relation
of GRBs to their macro-environments, and to compare the GRB-selected
galaxies to other high redshift samples. We perform both qualitative
and quantitative analyses by categorizing the galaxies according to
their visual properties, and by examining their surface brightness
profiles.  We find that all of the galaxies have approximately
exponential profiles, indicative of galactic disks, and have a median
scale length of about 1.7 kpc.  Inspection of the visual morphologies
reveals a high fraction of merging and interacting systems, with $\sim
30\%$ showing clear signs of interaction, and an additional $30\%$
exhibiting irregular and asymmetric structure which may be the result
of recent mergers; these fractions are independent of redshift and
galaxy luminosity.  On the other hand, the three GRB host galaxies for
which submillimeter and radio emission has been detected are isolated
and compact, unlike the luminous submillimeter-selected galaxies.  The
fraction of mergers appears to be elevated compared to other high
redshift samples, particularly for the low luminosities of GRB hosts
($M_B\sim -16$ to $-21$ mag).  This suggests that merging and
interacting galaxies undergoing a burst of star formation may be an
efficient site for the production of GRB progenitors.  Finally, we
show that GRB hosts clearly follow the size-luminosity relation
present in other galaxy samples, but thanks to absorption redshifts
they help extend this relation to lower luminosities.
\end{abstract}


\section{Introduction}
\label{sec:intro}

For nearly a century astronomers have attempted to classify galaxies
by their apparent shape and to draw conclusions about the process of
galaxy formation and evolution from these morphologies.  The basic
Hubble classification \citep{hub26} and its variants (e.g.,
\citealt{hub36,san61}) divide galaxies into three broad categories: 
elliptical, disk, and irregular.  This morphological classification
correlates with, among other physical properties, the star formation
activity of the galaxies.  Studies of local galaxy samples, as well as
the currently-favored cold dark matter ($\Lambda{\rm CDM}$) model,
also suggest that interactions and mergers play an important role in
the build up of galactic and stellar mass, through both the accretion
of material and an increase in the star formation rate.  Thus, the
morphological properties of galaxies as a function of cosmic time
provide direct insight into the physical processes governing galaxy
evolution.

In this manner, analysis of deep fields obtained primarily with the
\hst\ (HST) suggest that the locally-defined Hubble sequence may 
begin to break down at $z\sim 1$ \citep{van02} with the emergence of a
sizable fraction of faint, irregular, and interacting systems (e.g.,
\citealt{dwg95,ell97}).  While these observations, along with 
observations of local galaxies (e.g.,\citealt{arp66}), suggest that
mergers play a significant role in the formation of galaxies, three
important limitations prevent a conclusive connection between
morphology and galaxy formation at higher redshift.

First and foremost, studies of galaxy morphologies rely on
flux-limited samples, which may contain a large fraction of atypical
objects; e.g., ultra-luminous submillimeter galaxies
\cite{cwo+03}, or $R<25$ mag optically-selected galaxies
\cite{ccw03}.  Second, it is not clear how to relate the various
samples (e.g., Lyman break galaxies, submillimeter-selected galaxies,
near-IR selected galaxies) to the low redshift population or to each
other.  This is partly because of the different selection techniques
and the differences in observed properties and space densities.
Finally, at faint fluxes ($R\gtrsim 25$ mag), where irregular galaxies
may dominate the population, the distance scale relies on photometric
redshifts, whose accuracy is difficult to assess.

In this context, it is interesting to investigate the morphological
properties of $\gamma$-ray burst (GRB) host galaxies.  We now have
conclusive evidence that GRBs mark the death of massive stars
\cite{smg+03} and therefore pinpoint star-forming galaxies at all
redshifts \citep{hf99,bkd02,chg04}.  This allows a uniform selection
over a wide range of redshift and luminosity.  In addition, absorption
spectroscopy of the bright afterglows allows us to measure redshifts
of arbitrarily faint galaxies.  Thus, the current GRB host sample
spans $z\sim 0.1-4.5$ and $M_B\approx -16$ to $-21$ mag (i.e.,
$0.01L_*$ to $L_*$).

Here we present a comprehensive analysis of all optical HST
observations of GRB host galaxies.  The purpose of this study is
twofold: First, to obtain information on the large-scale environments
in which GRBs occur, as a clue to the formation of the progenitors.
Second, to survey a set of high redshift galaxies which are physically
related by their star formation activity, but which alleviate some of
the selection effects of other samples.  We summarize the HST
observations in \S\ref{sec:data}, provide a qualitative
(\S\ref{sec:morph}) and quantitative (\S\ref{sec:sb}) analysis of the
host morphologies, and compare the results to other high redshift
galaxy samples (\S\ref{sec:disc}).  We show that despite an overall
diversity in the sizes and luminosities of GRB hosts, they invariably
have roughly exponential profiles, with a large fraction undergoing
mergers and interactions.

\section{\hst\ Data}
\label{sec:data}

We retrieved data from the HST archive\footnotemark\footnotetext{\sf
http://archive.stsci.edu/hst/} for all available GRBs after
``on-the-fly'' pre-processing.  These include $29$ GRBs observed with
the Space Telescope Imaging Spectrograph (STIS), $8$ GRBs observed
with the Wide-Field Planetary Camera 2 (WFPC2), and $10$ GRBs observed
with the Advanced Camera for Surveys (ACS).  Details of the
observations are summarized in Table~\ref{tab:hst}.  For each GRB we
used the latest available images to reduce contamination from the
afterglow and/or supernova emission\footnotemark\footnotetext{Residual
afterglow and/or supernova emission is detected in GRBs 970228,
991216, 030329, and 041006.}.

We processed and combined individual exposures using the IRAF tasks
{\tt drizzle} (STIS, WFPC2) and {\tt multidrizzle} (ACS) in the {\tt
stsdas} package \citep{fh02}.  In all cases we used {\tt
pixfrac}$=0.8$, with {\tt pixscale}$=0.5$ for the STIS images, {\tt
pixscale}$=0.7$ for the WFPC2 images, and {\tt pixscale}$=1.0$ for the
ACS images.  The resulting images have pixel scales of 0.025, 0.07,
and 0.05 arcsec pix$^{-1}$, respectively.

\begin{figure}
\epsscale{1}
\plotone{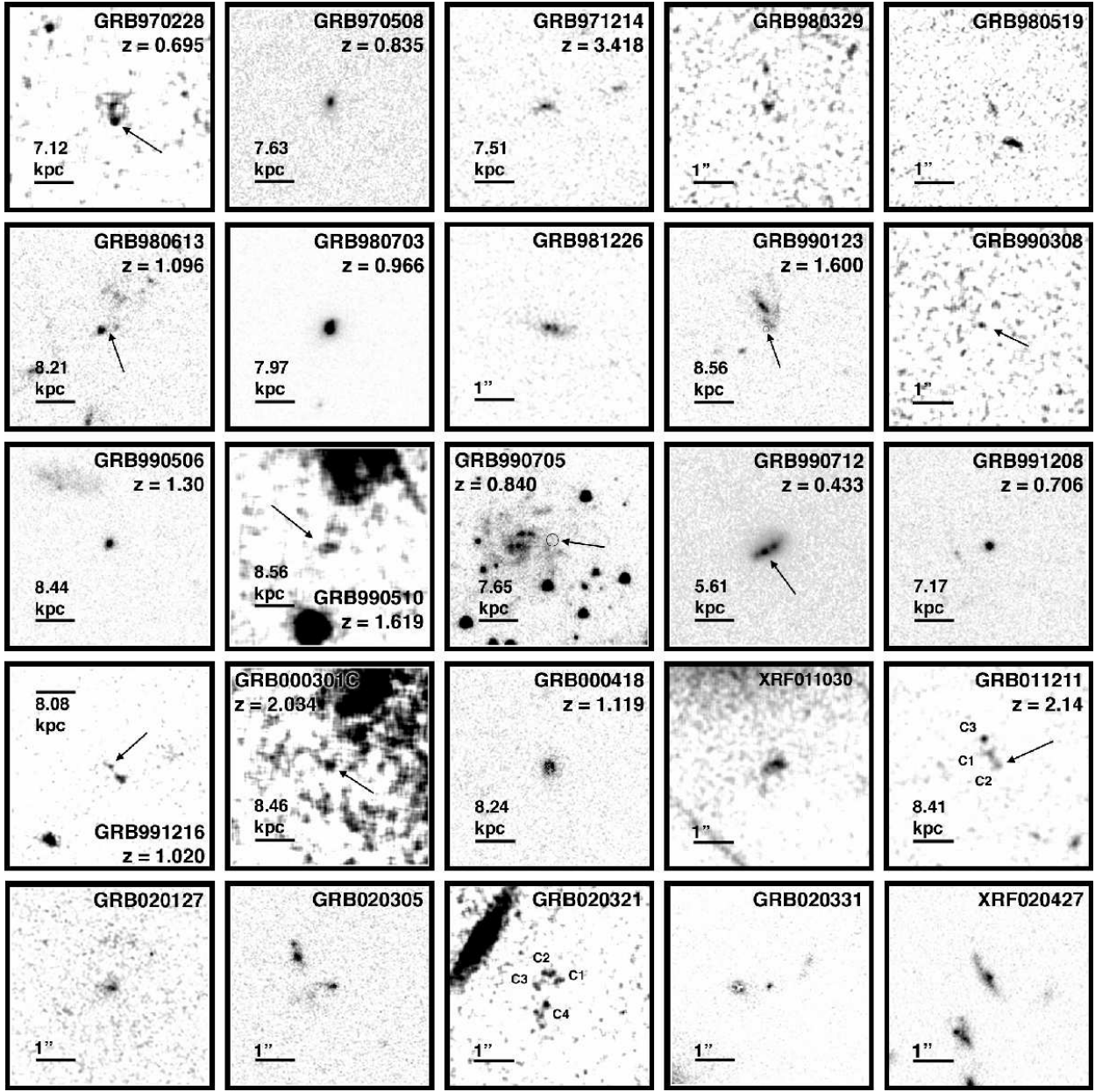}
\caption{\hst\ STIS images of GRB host galaxies.  Each panel is 
5\arcsec\ on a side and aligned such that north is up and east is to
the left.  Arrows mark the position of some host galaxies, or the
location of the GRB within complex systems.
\label{fig:stis}}
\end{figure}

\begin{figure}
\epsscale{0.9}
\plotone{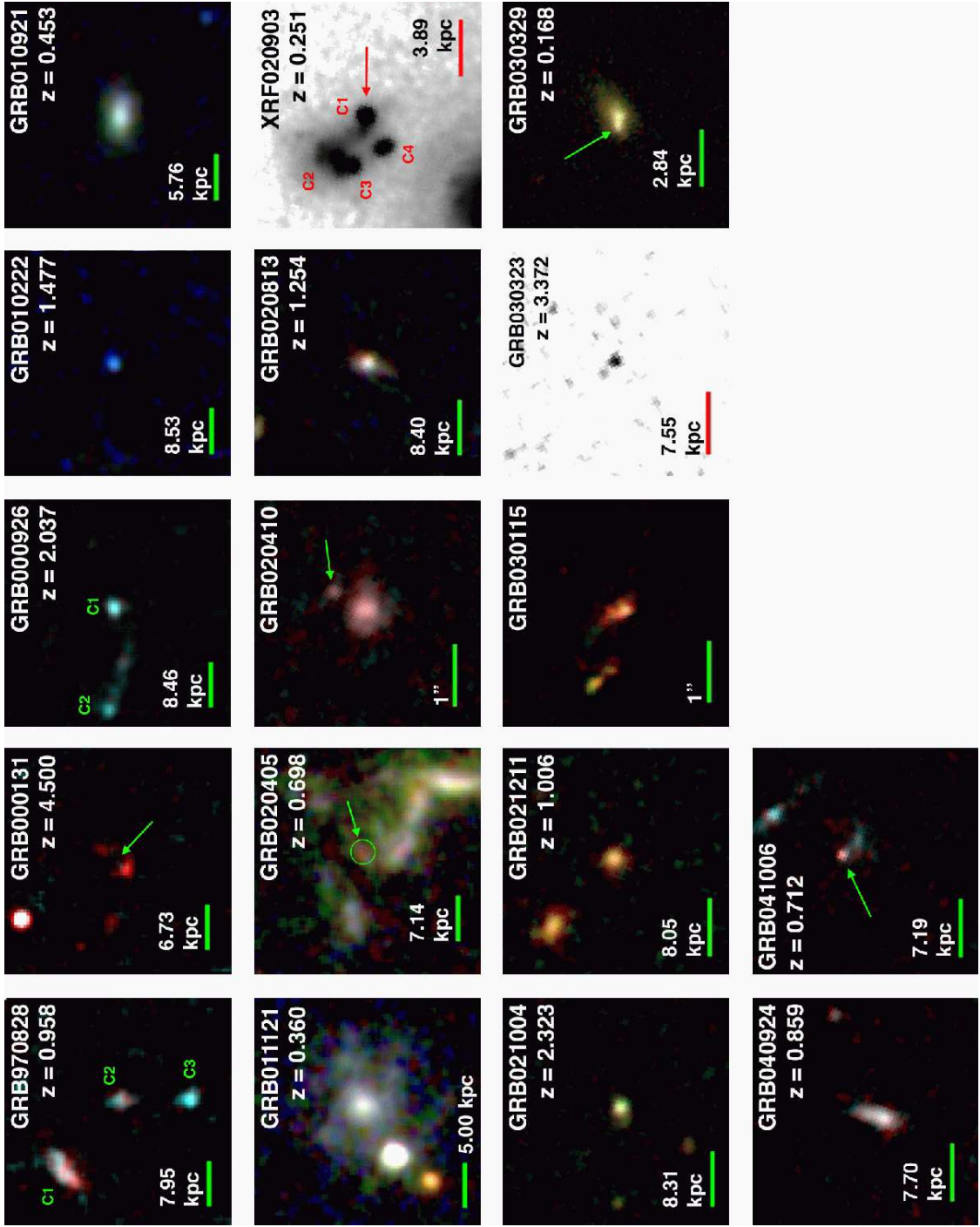}
\caption{\hst\ WFPC2 and ACS images and color composites of GRB host 
galaxies.  Each panel is 5\arcsec\ on a side and aligned such that
north is up and east is to the left.  Arrows mark the position of the
GRB within complex systems.  In the case of GRB\,041006 the supernova
which accompanied the burst \citep{} is visible as a red point source.
For GRB\,020410 the arrow marks the afterglow; it is not clear is the
bright galaxy to the south-east is the host galaxy, or if there is a
faint host underlying the afterglow position.
\label{fig:wfpc-acs}}
\end{figure}

In Figures~\ref{fig:stis} and \ref{fig:wfpc-acs} we show grayscale and
color images of the individual host galaxies.  All images are
flux-calibrated in the AB system according to the zero-points listed
in the instrument handbooks (see also \citealt{sjb+05}), and are
corrected for Galactic extinction \citep{sfd98}.  For GRB\,011121 we
used the extinction value determined by \citet{pbr+02} from
observations of the afterglow.

\section{Morphological Classification}
\label{sec:morph}

To classify the morphological properties of the GRB host galaxies, we
created eight different qualitative morphological categories.  Each
category is independent of the others and galaxies may fit into more
than one category.  In this manner we are able to place a large number
of galaxies into individual categories and account for multiple
features.  The categories are: concentrated elliptical or circular
structure, or blob-like (BL); conspicuous disk structure (D); highly
asymmetric or irregular structure (AS); galactic structure containing
knots (KN); galaxies with off-center peaks (OC); galaxies with tidal
tails (TT); galaxies that are either undergoing mergers or are closely
interacting (MI); and galaxies which are too faint for morphological
analysis (TF).  The classification has been carried out independently
by C.W.~and E.B., and the results are summarized in Tables
\ref{tab:hst} and \ref{tab:morph}.

Of the 45 host galaxies observed with HST, three are not detected in
our images, and an additional four are too faint to accurately
categorize.  These galaxies either occur at unknown redshifts, or
$z>1.5$.  While these galaxies cannot be classified morphologically,
they do indicate that a non-negligible fraction of GRBs (and hence of
the formation of massive stars) occurs in very low luminosity and/or
low surface brightness systems.  In fact, given the measured redshifts
and the magnitude limits for these galaxies we find that they
typically have an absolute rest-frame $B$-band magnitude of
$M_B\gtrsim -17.5$ mag, somewhat fainter than the Large Magellanic
Cloud.

For the remaining 38 galaxies with morphological classification we
combine the basic categories into two general groups: regular and
irregular/interacting.  Regular galaxies are those that are
categorized exclusively as either blob- or disk-like.  The primary
difference between these two categories is that the blob-like galaxies
have much higher luminosity concentrations, while the disk-like
galaxies have symmetric extended features.  In \S\ref{sec:sb} below we
show that both the BL and D galaxies have surface brightness
distributions that are well-described by exponential profiles and are
therefore disk galaxies.  The regular galaxies comprise about $30\%$
of the total sample.

The irregular category includes galaxies that are asymmetric or show
signs of a merger or interaction.  The latter includes multiple bright
galaxies (e.g., the hosts of GRB\,020405 and XRF\,020903), or galaxies
with filamentary structures (interpreted as tidal tails) extending
towards nearby galaxies with which they are interacting (e.g., the
host of GRB\,000926).  These tails are not symmetric about the center
of the light distribution.

A similar morphology is evident in the OC category, for which the
extended low surface brightness emission is not likely to be part of
an ordered disk structure.  The majority of these galaxies do not have
visible galactic neighbors.  If they are the results of galactic
mergers then they are most likely in the late stages of the merging
process, or alternatively signal an interaction with a lower mass
galaxy.  In this context the environment of GRB\,991208 may be of
particular interest.  A faint galaxy $\sim 7$ kpc from the host galaxy
exhibits a tidal tail morphology suggesting that the host is
interacting with a low mass companion.  It is also interesting to note
that the host of GRB\,991208, with $M_B\approx -18.2$ mag, is similar
in brightness to the LMC, suggesting that mergers between dwarfs play
an important role in the assembly of more massive galaxies; an
illustrative local example is NGC 1487 which is an interacting system
of two dwarf galaxies \citep{jb90}.

The asymmetric galaxies exhibit clumpiness or concavities in their
light distribution which may be interpreted as the result of an
interaction.  However, this morphology may also be interpreted as
clumpiness in the distribution of the star formation activity,
particularly in the case of higher redshift hosts for which we sample
the rest-frame UV light.  We note that the latter explanation may be
partially supported by the lack of obvious nearby companions, although
as in the case of the OC galaxies, these systems may be in the late
stages of merging.  The same argument holds for galaxies exhibiting
knots, which could be the remnant bright cores of merging galaxies or
signs of patchy star formation activity.  A relevant example is the
host of GRB\,990705.  This galaxy exhibits pronounced spiral structure
with bright knots of star formation.  At lower surface brightness (or
higher redshift) the spiral arms may not be detected and the system
might appear to have a knot morphology.

For the galaxies with morphological classification, the ratio of
irregular and merging or interacting systems to regular systems is
about $2:1$.  This ratio does not change significantly as a function
of redshift.  Dividing the sample into $z<1$ (low-$z$) and $z>1$
(high-$z$) bins, we find that at low-$z$ regular galaxies account for
about $36\%$ of the sample while $64\%$ are irregular.  For the
high-$z$ sample the fractions are $31\%$ and $56\%$, respectively,
with the remainder being too faint.  The only two categories with
possible evolution between the low- and high-$z$ samples are the tidal
tails and disks.  Tidal tails occur in about $7\%$ of the low-$z$
population, but appear in $25\%$ of the high-$z$ population, while
disk galaxies make up $36\%$ of the low-$z$ population and only $6\%$
of the high-$z$ population.  The latter trend may be attributed to
surface brightness dimming, but the increase in the frequency of tidal
features with redshift may be real since surface brightness dimming
would tend to have the opposite effect.

We finally note that some ambiguity exists amongst our broad
classifications.  For example, the host of GRB\,990123 may be
interpreted as a merger/interaction with strong tidal tails, where the
burst itself occurred in the disrupted galaxy.  Alternatively, this
galaxy may be classified as a disk galaxy with a bright spiral arm
accentuated by bright knots, which are presumably star forming HII
regions.  In this case, the burst was located in one of these bright
knots.  Still, such cases comprise a relatively small fraction of the
overall sample.

\section{Surface Brightness}
\label{sec:sb}

\begin{figure}
\epsscale{0.8}
\plotone{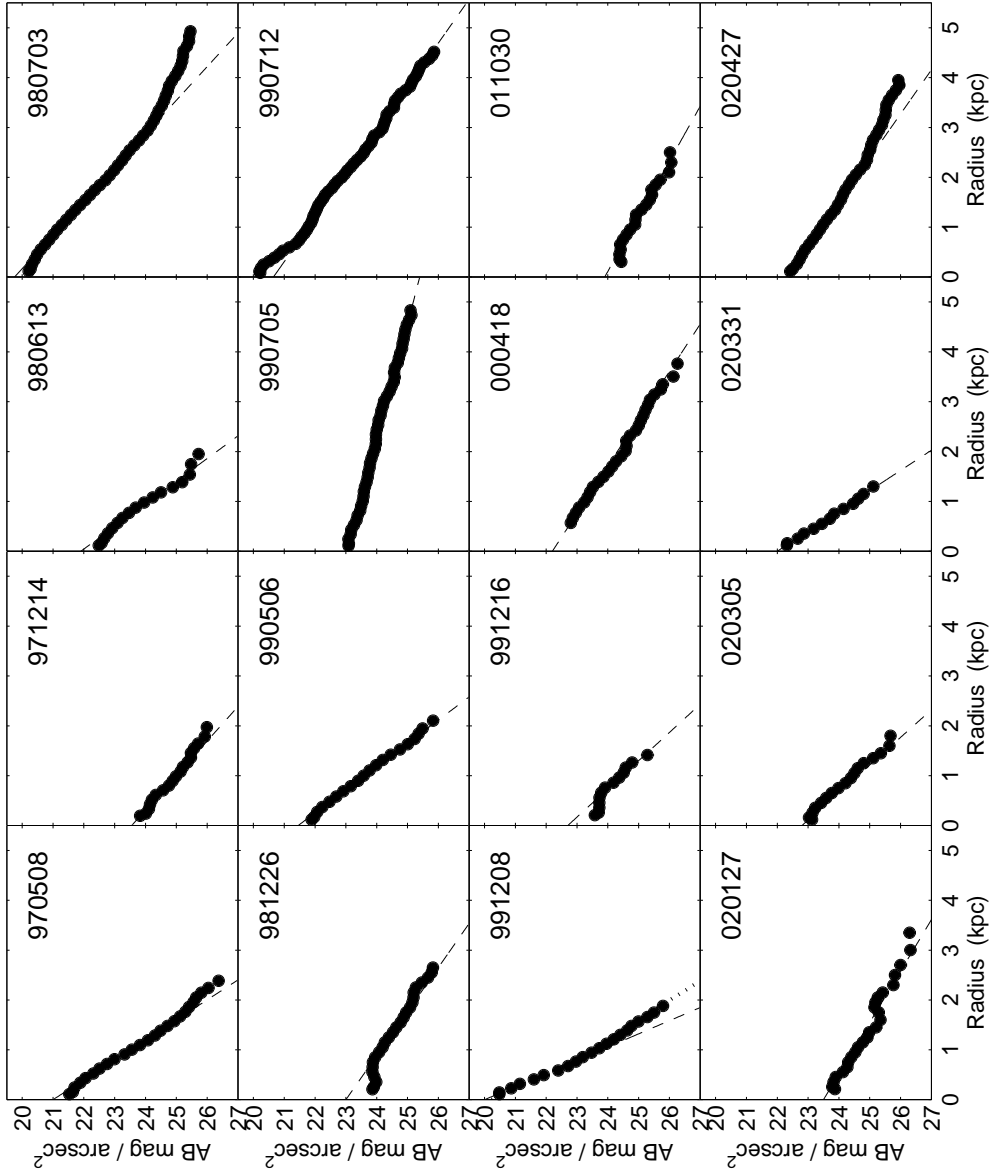}
\caption{Radial surface brightness profiles for GRB host galaxies 
observed with STIS.  The dotted line in the panel of GRB\,991208 is
the instrumental point spread function of STIS as measured from
several stars in the field.  While the host of GRB\,991208 is
consistent with a point source, all the other host galaxies are well
resolved.  The dashed lines are exponential disk fits to the data.
\label{fig:sbstis}}
\end{figure}

\begin{figure}
\epsscale{0.7}
\plotone{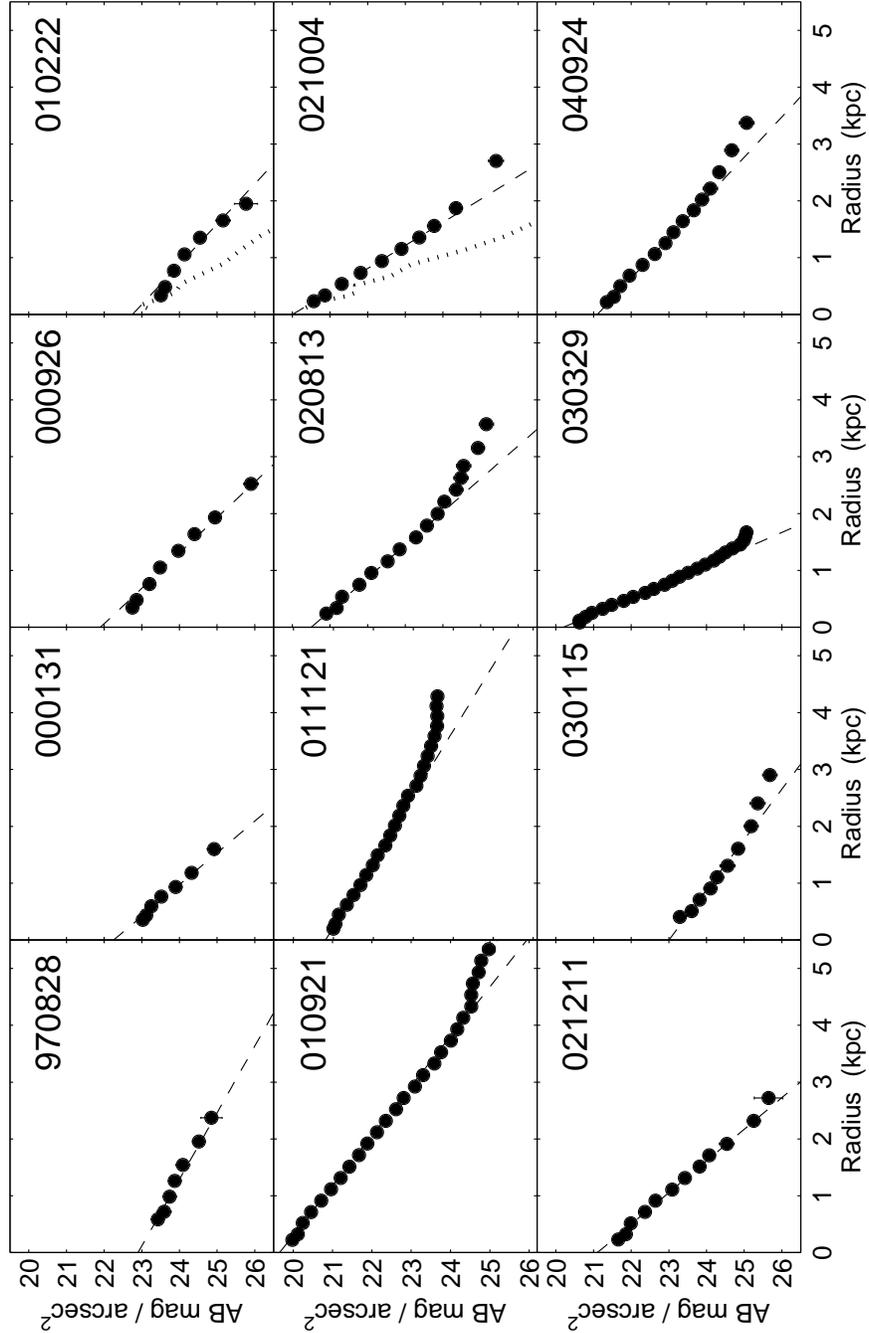}
\caption{Radial surface brightness profiles for GRB host galaxies 
observed with WFPC2 and ACS.  The dotted line in the panel of GRBs
010222 and 021004 are the instrumental point spread function of WFPC2
and ACS, respectively as measured from several stars in the field.
Clearly, all of the GRB host galaxies are well resolved.  The dashed
lines are exponential disk fits to the data.
\label{fig:sbacs}}
\end{figure}

\begin{figure}
\epsscale{0.9}
\plotone{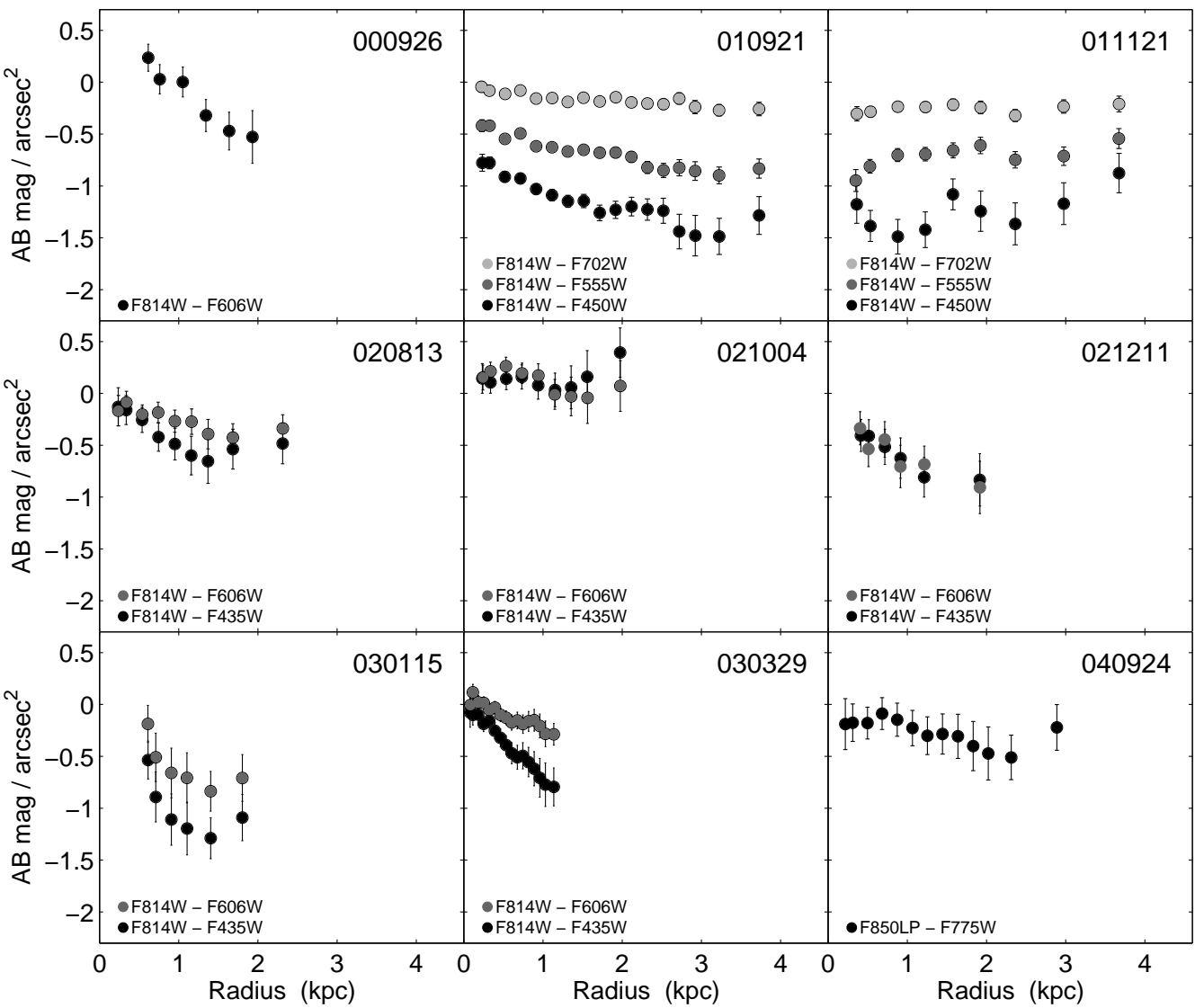}
\caption{Radial surface brightness colors for GRB host galaxies 
observed with WFPC2 and ACS.
\label{fig:sbcolor}}
\end{figure}

Observations of local galaxies suggest that the surface brightness
profiles of disk galaxies are roughly exponential, while those of
elliptical galaxies and galaxy bulges tend to follow an $r^{1/4}$ de
Vaucouleurs law \citep{vaucouleurs48}.  In this section we determine
the surface brightness profiles and sizes of the GRB host galaxies and
investigate their distributions as an additional input into their
morphological classification.  We determine the radial surface
brightness distributions in two ways.  First, for galaxies with a
relatively simple apparent morphology we construct radial surface
brightness plots using the IRAF task {\tt phot}, with a range of
apertures chosen to span the full extent of each host galaxy while
maintaining $S/N\gtrsim 5$ in each bin.  A comparison of circular
apertures to elliptical isophotes (using the IRAF task {\tt ellipse})
indicates that the difference is typically within the uncertainty in
individual apertures.  The resulting surface brightness profiles are
shown in Figures~\ref{fig:sbstis} and \ref{fig:sbacs}.  For the host
galaxies observed with WFPC2 and ACS in multiple filters we also plot
the radial color profiles (Figure~\ref{fig:sbcolor}).

None of the well-resolved host galaxies with high signal-to-noise
detections exhibit a clear $r^{1/4}$ profile, confirming their nature
as disk and irregular galaxies.  With the exception of the host galaxy
of GRB\,991208 all the galaxies are well-resolved relative to the
instrumental point spread function as measured from stars in the field
(Figures~\ref{fig:sbstis} and \ref{fig:sbacs}).  We thus fit the
surface brightness profiles of all systems with an exponential disk:
$\Sigma(r)=\Sigma_0{\rm exp}(-r/r_s)$, leaving the central surface
brightness ($\Sigma_0$) and the scale length ($r_s$) as free
parameters.  We find that the scale length distribution peaks at
$r_s\approx 0.09\arcsec$, with a tail extending to $\sim 0.35\arcsec$.

For the host galaxies observed in multiple filters we find that about
$60\%$ exhibit a color gradient as a function of radius
(Figure~\ref{fig:sbcolor}), becoming redder at large radii. The single
exception to this trend is the host galaxy of GRB\,011121, which is
somewhat bluer at larger radii.  Since blue light traces recent star
formation, the observed trend suggests that the star formation
activity in GRB host galaxies is more concentrated than the overall
light distribution.  The trend observed for GRB\,011121 may suggest
that substantial star formation is taking place across the whole disk
of the galaxy.  While color information is not available for the host
of GRB\,990705, it too has fairly distributed star formation activity
as shown by its spiral arm structure and bright knots.

Our second approach in studying the surface brightness profiles is to
use the GALFIT software \citep{phi+02}.  This allows us to fit all but the
most irregularly shaped galaxy (GRB\,020405).  In this case we use the
generalized Sersic function
\citep{sersic68}
\begin{equation}
\Sigma(r) = \Sigma_e{\rm exp}[-\kappa((r/r_e)^{1/n}-1)],
\end{equation}
where $n$ is the concentration parameter ($n=1$ is equivalent to an
exponential disk, while $n=4$ is the de Vaucouleurs profile), $\kappa$
is a constant that is coupled to the value of $n$, $r_e$ is the
effective radius, and $\Sigma_e$ is the surface
brightness\footnotemark\footnotetext{The length scale of the
exponential disk defined above is related to the effective radius as
$r_s=r_e/1.68$, while $\Sigma_0=5.36\Sigma_e$.} at $r=r_e$.  We
generated point spread functions for the individual instruments and
filters using the Tiny Tim software
package\footnotemark\footnotetext{\sf
http://www.stsci.edu/software/tinytim/tinytim.html}, assuming a power
law spectrum $F_\nu\propto\nu^{-1}$, which is roughly appropriate for
the observed color distribution of GRB host galaxies \citep{bck+03}.
In all cases we find adequate fits to the host galaxies, with
$\chi^2_r\approx 0.5-2$ per degree of freedom.  We note that some
sources, particularly at low signal-to-noise, can be adequately fit
with a range of $n\sim 1-4$.  For sources with complex morphology
(e.g., XRF\,020903) or contaminating point sources (e.g., GRB\,011121)
we use multiple components to account for substructure.  The resulting
values of $n$ and $r_e$ are listed in Table~\ref{tab:hst}.

\begin{figure}
\epsscale{0.65}
\plotone{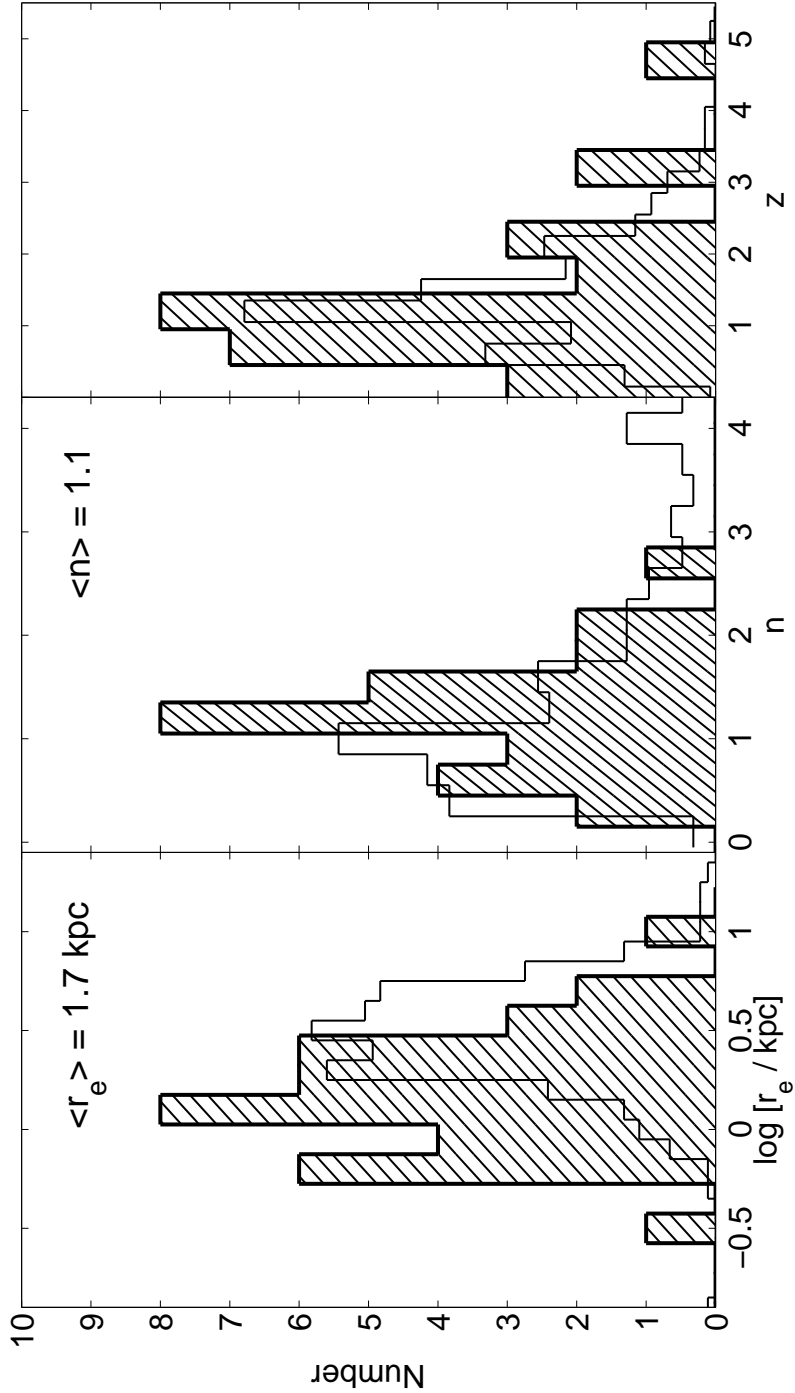}
\caption{Histograms of the effective radius ($r_e$), the Sersic 
profile parameter $n$, and the redshifts for GRB host galaxies
(hatched), and for galaxies from the FIRES survey \citep{} (thin
line).  The distribution of $n$ is sharply peaked around a value of 1
suggesting that GRB host galaxies are well described by exponential
disks.  In addition, GRB host galaxies are on average a factor of two
smaller compared to the FIRES galaxies.
\label{fig:hists}}
\end{figure}

In Figure~\ref{fig:hists} we plot the distribution of $r_e$ and $n$
for the hosts with an accurate value of $n$.  For hosts without a
known redshift we take advantage of the flat evolution of the angular
diameter distance with redshift, and assume a value of $8$ kpc
arcsec$^{-1}$ appropriate for $z\sim 1-3$.  The distribution of $n$ is
strongly peaked around a value of $\sim 1$, indicating that GRB hosts
are well described as exponential disks.  As noted in other studies
(e.g., \citealt{rfc+04}), $n\lesssim 2$ is an efficient criterion for
disk-dominated galaxies.  The distribution of $r_e$ ranges from about
0.3 to 10 kpc, with a peak at $r_e\approx 1.7$ kpc.  As shown in
Figure~\ref{fig:renzmb} we do not find any correlation between $r_e$
and $n$ or redshift, although we note that there is a larger
dispersion in $r_e$ for $z\lesssim 1$.  This may be a result of
surface brightness dimming which would tend to make higher redshift
objects appear more compact.

A comparison to the morphological analysis of galaxies in the FIRES
data \citep{trr+04,tfr+05} suggests that the distributions of $n$
values are similar, with the exception that the latter exhibit a tail
at $n\gtrsim 3$ (ellipticals) which may not present in the GRB sample.
The distribution of effective radii, however, peaks at a large value
compared to the GRB sample.  To provide a direct comparison we
corrected the values given in \citet{trr+04} and \citet{tfr+05} for
ellipticity and for the systematic over-estimate of about $15\%$
compared to GALFIT results (see Figure 4 of \citealt{trr+04}).  The
median effective radius of the FIRES galaxies is about a factor of two
higher than that of the GRB sample.

The underlying reason for the smaller sizes of GRB host galaxies is
revealed in the correlation between the effective radius and the
rest-frame absolute $B$-band magnitude, $M_B$
(Figure~\ref{fig:renzmb}).  The slope of the correlation for GRB hosts
is remarkably similar to the relation found by \citet{fre70} for local
exponential disks, but with a surface brightness that is about $1-1.5$
mag arcsec$^{-2}$ higher.  This is similar to the results found from
the FIRES data, with the exception that the FIRES galaxies are
brighter (and hence larger).

\section{Discussion}
\label{sec:disc}

The sample of 42 GRB host galaxies imaged with HST yields several
interesting trends.  The host morphologies and surface brightness
profiles indicate that GRB hosts are well described as exponential
disks with sizes ranging from $\sim 0.5-5$ kpc.  In addition, GRB
hosts exhibit a large fraction of interactions or mergers,
particularly when compared to galaxies of similar luminosity from
other surveys \citep{ccw03}.  With the exception of GRB\,990705, the
GRB hosts lack distinctive spiral structure despite having
predominantly disk dominated surface brightness profiles.

As shown in \S\ref{sec:sb} the bulk of the galaxies have exponential
surface brightness profiles.  There are a few minor exceptions in low
signal-to-noise filters, as well as in the case of GRB\,021004, for
which we find an adequate fit with an $r^{1/4}$ profile using GALFIT;
a fit with $n=1$ is equally adequate.  Except for GRB\,990705, and
possibly GRB\,990123, none of the GRB host galaxies exhibit clear
signs of spiral structure.  This includes in particular the several
hosts at $z<0.5$, for which such structures should be easily detected.
\citet{cgj+04} show that spiral and bar structures should be visible
at redshifts as high as $z\sim 2.3$.  The lack of ordered spiral
structure in GRB hosts may point to a violent merger history which
suppresses the emergence of spiral arms.

\begin{figure}
\epsscale{0.6}
\plotone{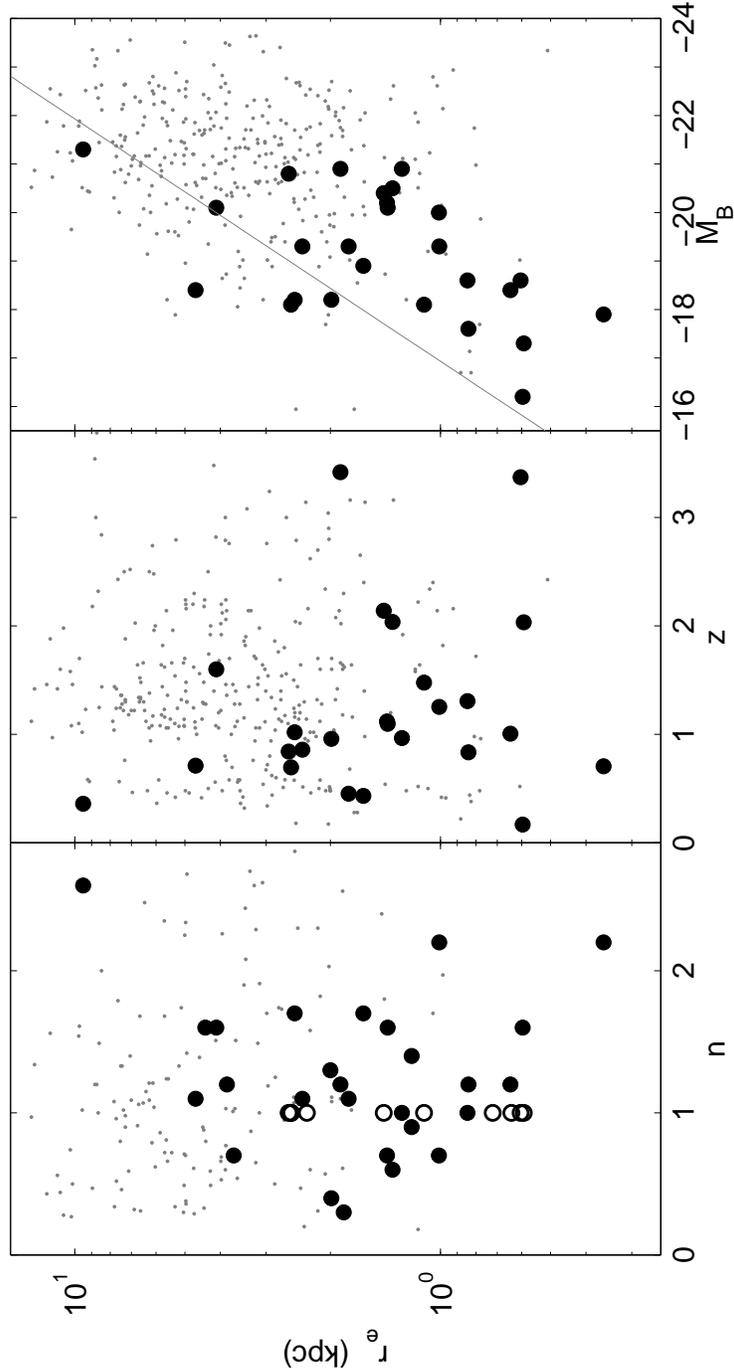}
\caption{Effective radius plotted against $n$, redshift, and absolute 
$B$-band rest-frame magnitude for GRB hosts (black circles) and FIRES
galaxies (gray dots).  No clear trend is evident between $r_e$ and $n$
or $z$, but there is a larger dispersion in $r_e$ at $z\lesssim 1$ for
the GRB hosts, which is possibly missing at higher redshift due to
surface brightness dimming.  There is a clear trend between $r_e$ and
$M_B$ which is similar to the Freeman relation for local exponential
disks (gray line), but with a surface brightness that is higher by
about $1-1.5$ mag arcsec$^{-2}$.  The GRB hosts extend the
size-luminosity relation to lower luminosities than the FIRES sample.
\label{fig:renzmb}}
\end{figure}

The observed size-luminosity correlation presented in
Figure~\ref{fig:renzmb} is in good agreement with that observed in
other galaxy samples (e.g., \citealt{tfr+05}).  However, the GRB
sample extends this relation to lower luminosities due to the
availability of absorption redshifts which are not subject to the
brightness limit for spectroscopy imposed on flux-limited surveys.

Three of the galaxies in our sample, GRBs 980703, 000418, and 010222,
exhibit high luminosity at submillimeter and/or radio wavelengths
\citep{bkf01,fbm+02,bck+03}.  Contrary to the trend observed in
submillimeter-selected galaxies
\citep{cwo+03,ccw03}, all three are highly symmetric and show no 
clear signs of interaction.  Each was categorized as a blob-like
galaxy, while GRB\,980703 was additionally categorized as off-center.
However, it has the most modest deviation of any of the off-center
galaxies in our sample.  In comparison, HST observations of
submillimeter galaxies indicate a merger fraction of $\sim 0.4-0.8$
\citep{ccw03} and a low percentage ($\lesssim 20\%$) of
symmetrically shaped galaxies \citep{cwo+03}.  These three galaxies
are also significantly bluer than the typical submillimeter-selected
galaxies \citep{bck+03}.  Taken together, these properties suggest
that the submillimeter and radio bright GRB hosts are a distinct
population from the field submillimeter-selected galaxies.

However, on the whole, a large fraction of the GRB host galaxies show
evidence of merging or interaction.  About $30-60\%$ of our galaxies
show evidence of merging.  This proportion appears to be independent
of both redshift and apparent magnitude.  In other high redshift
surveys, the proportion of interacting galaxies increases with galaxy
brightness \citep{ccw03}.  This discrepancy may be caused by one of
several factors.  First, it is possible that previous surveys have a
selection bias against faint merging galaxies due to a flux limit.  If
this is the case, then previous surveys under-represent an important
category of galaxy morphology.  Alternatively, the general correlation
between magnitude and frequency of merging galaxies could be correct,
and some physical process causes GRBs to preferentially occur in faint
merging systems, instead of bright ones, for example a preference for
low metallicity \citep{}.  Overall, the high fraction of galaxies
which show signs of merging and interaction indicate that these are
regions of elevated star formation activity, and that GRBs are less
likely to occur in stable disk galaxies.

\acknowledgements 
We thank Alicia Soderberg, Francois Schweizer, Leonidas Moustakas,
Marla Geha, and Swara Ravindranath for useful comments and
discussions.  E.B. is supported by NASA through Hubble Fellowship
grant HST-01171.01 awarded by the Space Telescope Science Institute,
which is operated by AURA, Inc., for NASA under contract NAS5-26555.
Additional support was provided by an archival NASA grant.


\begin{thebibliography}{}
 
\bibitem[\protect\citeauthoryear{{Arp}}{{Arp}}{1966}]{arp66}
{Arp}, H. 1966, \apjs, 14, 1
 
\bibitem[\protect\citeauthoryear{{Berger} et~al.}{{Berger}
  et~al.}{2003}]{bck+03}
{Berger}, E., {Cowie}, L.~L., {Kulkarni}, S.~R., {Frail}, D.~A., {Aussel}, H.,
  \& {Barger}, A.~J. 2003, \apj, 588, 99
 
\bibitem[\protect\citeauthoryear{{Berger}, {Kulkarni}, \& {Frail}}{{Berger}
  et~al.}{2001}]{bkf01}
{Berger}, E., {Kulkarni}, S.~R.,  \& {Frail}, D.~A. 2001, \apj, 560, 652
 
\bibitem[\protect\citeauthoryear{{Bloom}, {Kulkarni}, \& {Djorgovski}}{{Bloom}
  et~al.}{2002}]{bkd02}
{Bloom}, J.~S., {Kulkarni}, S.~R.,  \& {Djorgovski}, S.~G. 2002, \aj, 123, 1111
 
\bibitem[\protect\citeauthoryear{{Chapman} et~al.}{{Chapman}
  et~al.}{2003}]{cwo+03}
{Chapman}, S.~C., {Windhorst}, R., {Odewahn}, S., {Yan}, H.,  \& {Conselice},
  C. 2003, \apj, 599, 92
                                                                                                                                                 
\bibitem[\protect\citeauthoryear{{Christensen}, {Hjorth}, \&
  {Gorosabel}}{{Christensen} et~al.}{2004}]{chg04}
{Christensen}, L., {Hjorth}, J.,  \& {Gorosabel}, J. 2004, \aap, 425, 913
 
\bibitem[\protect\citeauthoryear{{Conselice}, {Chapman}, \&
  {Windhorst}}{{Conselice} et~al.}{2003}]{ccw03}
{Conselice}, C.~J., {Chapman}, S.~C.,  \& {Windhorst}, R.~A. 2003, \apjl, 596,
  L5
 
\bibitem[\protect\citeauthoryear{{Conselice} et~al.}{{Conselice}
  et~al.}{2004}]{cgj+04}
{Conselice}, C.~J., et~al. 2004, \apjl, 600, L139
 
\bibitem[\protect\citeauthoryear{{de Vaucouleurs}}{{de
  Vaucouleurs}}{1948}]{vaucouleurs48}
{de Vaucouleurs}, G. 1948, Annales d'Astrophysique, 11, 247
 
\bibitem[\protect\citeauthoryear{{Driver}, {Windhorst}, \&
  {Griffiths}}{{Driver} et~al.}{1995}]{dwg95}
{Driver}, S.~P., {Windhorst}, R.~A.,  \& {Griffiths}, R.~E. 1995, \apj, 453, 48
 
\bibitem[\protect\citeauthoryear{{Ellis}}{{Ellis}}{1997}]{ell97}
{Ellis}, R.~S. 1997, \araa, 35, 389
 
\bibitem[\protect\citeauthoryear{{Frail} et~al.}{{Frail} et~al.}{2002}]{fbm+02}
{Frail}, D.~A., et~al. 2002, \apj, 565, 829
 
\bibitem[\protect\citeauthoryear{{Freeman}}{{Freeman}}{1970}]{fre70}
{Freeman}, K.~C. 1970, \apj, 160, 811
 
\bibitem[\protect\citeauthoryear{{Fruchter} \& {Hook}}{{Fruchter} \&
  {Hook}}{2002}]{fh02}
{Fruchter}, A.~S.,  \& {Hook}, R.~N. 2002, \pasp, 114, 144
 
\bibitem[\protect\citeauthoryear{{Hogg} \& {Fruchter}}{{Hogg} \&
  {Fruchter}}{1999}]{hf99}
{Hogg}, D.~W.,  \& {Fruchter}, A.~S. 1999, \apj, 520, 54
 
\bibitem[\protect\citeauthoryear{{Hubble}}{{Hubble}}{1926}]{hub26}
{Hubble}, E.~P. 1926, \apj, 64, 321
 
\bibitem[\protect\citeauthoryear{{Hubble}}{{Hubble}}{1936}]{hub36}
{Hubble}, E.~P. 1936, Yale University Press
                                                                                                                                                 
\bibitem[\protect\citeauthoryear{{Johansson} \& {Bergvall}}{{Johansson} \&
  {Bergvall}}{1990}]{jb90}
{Johansson}, L.,  \& {Bergvall}, N. 1990, \aaps, 86, 167
 
\bibitem[\protect\citeauthoryear{{Peng} et~al.}{{Peng} et~al.}{2002}]{phi+02}
{Peng}, C.~Y., {Ho}, L.~C., {Impey}, C.~D.,  \& {Rix}, H. 2002, \aj, 124, 266
 
\bibitem[\protect\citeauthoryear{{Price} et~al.}{{Price} et~al.}{2002}]{pbr+02}
{Price}, P.~A., et~al. 2002, \apjl, 572, L51
 
\bibitem[\protect\citeauthoryear{{Ravindranath} et~al.}{{Ravindranath}
  et~al.}{2004}]{rfc+04}
{Ravindranath}, S., et~al. 2004, \apjl, 604, L9
 
\bibitem[\protect\citeauthoryear{{Sandage}}{{Sandage}}{1961}]{san61}
{Sandage}, A. 1961, {The Hubble atlas of galaxies} (Washington: Carnegie
  Institution, 1961)
 
\bibitem[\protect\citeauthoryear{{Schlegel}, {Finkbeiner}, \&
  {Davis}}{{Schlegel} et~al.}{1998}]{sfd98}
{Schlegel}, D.~J., {Finkbeiner}, D.~P.,  \& {Davis}, M. 1998, \apj, 500, 525
                                                                                                                                                 
\bibitem[\protect\citeauthoryear{{Sersic}}{{Sersic}}{1968}]{sersic68}
{Sersic}, J.~L. 1968, {Atlas de galaxias australes} (Cordoba, Argentina:
  Observatorio Astronomico, 1968)
 
\bibitem[\protect\citeauthoryear{Sirianni et~al.}{Sirianni
  et~al.}{2005}]{sjb+05}
Sirianni, M., et~al. 2005, {PASP accepted}
 
\bibitem[\protect\citeauthoryear{{Stanek} et~al.}{{Stanek}
  et~al.}{2003}]{smg+03}
{Stanek}, K.~Z., et~al. 2003, \apjl, 591, L17
 
\bibitem[\protect\citeauthoryear{Trujillo et~al.}{Trujillo
  et~al.}{2005}]{tfr+05}
Trujillo, I., et~al. 2005, {astro-ph/0504225}
 
\bibitem[\protect\citeauthoryear{{Trujillo} et~al.}{{Trujillo}
  et~al.}{2004}]{trr+04}
{Trujillo}, I., et~al. 2004, \apj, 604, 521
 
\bibitem[\protect\citeauthoryear{{van den Bergh}}{{van den
  Bergh}}{2002}]{van02}
{van den Bergh}, S. 2002, \pasp, 114, 797
 
\end{thebibliography}

\clearpage
\begin{deluxetable}{rcllclrcccl}
\rotate
\tabletypesize{\scriptsize}
\tablecolumns{11}
\tabcolsep0.08in\footnotesize
\tablewidth{0pc}
\tablecaption{HST Observations and Morphological Properties of GRB Host Galaxies 
\label{tab:hst}}
\tablehead {
\colhead {GRB}       		&
\colhead {$z$}		        &
\colhead {Instrument} 		&
\colhead {Filter}  		&
\colhead {Exp.~time}     	&
\colhead {Date}  		&
\colhead {$n$}   		&
\colhead {$r_e$} 		&
\colhead {$r_e$}	 	&
\colhead {AB Mag.}    		&
\colhead {Classification}       \\
\colhead {} 	  		    &
\colhead {}	 	            &
\colhead {}       		    &      
\colhead {}		       	    &      
\colhead {(s)}		 	    &
\colhead {}   			    &
\colhead {} 			    &
\colhead {(arcsec)}       	    &       
\colhead {(kpc)}       		    &      
\colhead {}	 		    &
\colhead {}
}
\startdata
970228  & 0.695 & STIS  & CL    & 2300  & 1997 Sep 4  & 1    & 0.36 & 2.53 & 24.7 & D \\
970508  & 0.835 & STIS  & CL    & 11568 & 1998 Aug 5  & 1.2  & 0.11 & 0.81 & 25.0 & BL    \\
970828  & 0.958 & WFPC2 & F606W & 3300  & 2001 Aug 16 &      &      &      &      & AS, MI    \\
C1      &       &       &       &       &             & 0.6  & 0.46 & 3.66 & 24.9 &  \\
C2	&	&	&	&       &	      & 0.4  & 0.25 & 1.99 & 25.9 &  \\
C3      &       &       &       &       &             & 0.6  & 0.16 & 1.27 & 25.8 &  \\
970828  &       & WFPC2 & F814W & 3300  & 2001 Aug 18 &      &      &      &      &  \\
C1      &       &       &       &       &             & 0.7  & 0.54 & 4.29 & 23.6 &  \\
C2      &       &       &       &       &             & 2.1  & 0.53 & 4.21 & 24.4 &  \\
C3      &       &       &       &       &             & 0.5  & 0.28 & 2.23 & 25.4 &  \\
971214  & 3.418 & STIS  & CL    & 8540  & 2000 Jun 12 & 1.2  & 0.25 & 1.91 & 26.3 & AS    \\
980326  & \nod  & STIS  & CL    & 7080  & 2000 Dec 31 & \multicolumn{5}{c}{not detected} \\
980329  & \nod  & STIS  & CL    & 8012  & 2000 Aug 24 & 0.9  & 0.15 & \nod & 27.0 & AS, MI, TT \\
980519  & \nod  & STIS  & CL    & 8924  & 2000 Jun 7  & 1    & 0.32 & \nod & 27.4 & TF    \\
980613  & 1.097 & STIS  & CL    & 5792  & 2000 Aug 20 & 1.6  & 0.17 & 1.42 & 25.1 & AS, KN, MI \\
980703  & 0.966 & STIS  & CL    & 5118  & 2000 Jun 18 & 1.0  & 0.16 & 1.29 & 22.7 & BL, OC \\
981226  & \nod  & STIS  & CL    & 7805  & 2000 Jul 3  & 1.2  & 0.48 & \nod & 25.2 & D, KN, MI  \\
990123  & 1.600 & STIS  & CL    & 5280  & 2000 Feb 7  &      &      &      & 24.4 & D, KN, MI, TT \\
knot    &       &       &       &       &             & 1    & 0.18 & 1.58 &      &	    \\
main    &       &       &       &       &             & 1.6  & 0.48 & 4.14 &      &	    \\
990308  & \nod  & STIS  & CL    & 7782  & 2000 Jun 19 & 1    & 0.09 & \nod & 28.8 & TF    \\
990506  & 1.307 & STIS  & CL    & 7856  & 2000 Jun 24 & 1.0  & 0.10 & 0.86 & 25.3 & BL    \\
990510  & 1.619 & STIS  & CL    & 5840  & 2000 Apr 29 & 1    & 0.08 & 0.72 & 27.7 & TF    \\
990705  & 0.842 & STIS  & CL    & 8792  & 2000 Jul 25 &      &      &      & 22.6 & D, KN  \\
bukge   &       &       &       &       &             & 4.0  & 0.10 & 0.78 &      &	    \\
disk    &       &       &       &       &             & 1    & 0.94 & 7.21 &      &	    \\
990712  & 0.433 & STIS  & CL    & 3720  & 2000 Apr 24 & 1.7  & 0.29 & 1.62 & 22.3 & D, KN, MI \\
C1      &       &       &       &       &             & 0.6  & 0.29 & 1.62 &      &	    \\
C2      &       &       &       &       &             & 1.0  & 0.20 & 1.15 &      &	    \\
991208  & 0.706 & STIS  & CL    & 3840  & 2000 Aug 3  & 2.2  & 0.05 & 0.38 & 24.5 & BL, MI?  \\
991216  & 1.020 & STIS  & CL    & 4720  & 2000 Apr 17 & 1.7  & 0.31 & 2.48 & 23.6 & AS, TT \\
000131  & 4.500 & WFPC2 & F606W & 8800  & 2001 Aug 20 & \nod & \nod & \nod & 28.0 & AS, MI \\
000131  &       & WFPC2 & F814W & 8800  & 2001 Aug 17 &      &      &      & 25.7 &       \\
C1	&	&	&	&	&	      & 0.7  & 0.15 & 1.00 &      & \\
C2	&	&	&	&	&	      & 1.0  & 0.41 & 2.75 &	  & \\
000301C & 2.034 & STIS  & CL    & 7031  & 2001 Feb 25 & 1    & 0.07 & 0.58 & 30.0 & TF    \\
000418  & 1.119 & STIS  & CL    & 5120  & 2001 Feb 11 & 0.7  & 0.17 & 1.38 & 24.7 & BL    \\
000926  & 2.037 & WFPC2 & F606W & 4400  & 2001 May 19 & 0.8  & 0.13 & 1.11 & 25.2 & MI, TT    \\
000926  &       & WFPC2 & F814W & 4400  & 2001 May 20 & 0.4  & 0.19 & 1.57 & 25.2 &       \\
010222  & 1.477 & WFPC2 & F450W & 6000  & 2001 Sep 8  & 1    & 0.14 & 1.19 & 26.0 & BL    \\
010222  &       & WFPC2 & F606W & 6000  & 2001 Sep 8  & 1    & 0.11 & 0.95 & 26.2 &       \\
010222  &       & WFPC2 & F814W & 6000  & 2001 Sep 9  & 1    & 0.15 & 1.31 & 26.2 &       \\
010921  & 0.453 & WFPC2 & F450W & 4400  & 2001 Dec 21 & 1.1  & 0.27 & 1.55 & 22.6 & D    \\
010921  &       & WFPC2 & F555W & 4400  & 2001 Dec 21 & 1.1  & 0.29 & 1.64 & 22.1 &       \\
010921  &       & WFPC2 & F702W & 4400  & 2001 Dec 22 & 1.2  & 0.31 & 1.80 & 21.6 &       \\
010921  &       & WFPC2 & F814W & 4400  & 2001 Dec 22 & 1.1  & 0.33 & 1.88 & 21.4 &       \\
010921  &       & WFPC2 & F850L & 4400  & 2001 Dec 22 & 1.0  & 0.31 & 1.80 & 21.3 &       \\
011030  & \nod  & STIS  & CL    & 7505  & 2002 Jun 12 & 0.3  & 0.23 & \nod & 24.9 & AS, OC    \\
011121  & 0.360 & WFPC2 & F450W & 4500  & 2002 Apr 21 & 2.6  & 6.48 & 32.4 & 24.1 & D    \\
011121  &       & WFPC2 & F555W & 4500  & 2002 May 2  & 2.6  & 3.13 & 15.6 & 23.4 &       \\
011121  &       & WFPC2 & F702W & 4500  & 2002 Apr 29 & 2.7  & 2.51 & 12.5 & 22.5 &       \\
011121  &       & WFPC2 & F814W & 4500  & 2002 Apr 29 & 2.6  & 1.88 & 9.4  & 22.2 &       \\
011121  &       & WFPC2 & F850L & 4500  & 2002 May 2  & 2.6  & 1.88 & 9.4  & 21.9 &       \\
011211  & 2.140 & STIS  & CL    & 4721  & 2002 Feb 9  &      &      &      & 26.6 & KN, MI    \\
C1      &       &       &       &       &             & 1    & 0.17 & 1.43 &      &       \\ 
C2      &       &       &       &       &             & 1    & 0.10 & 0.84 &      &       \\ 
C3      &       &       &       &       &             & 1    & 0.06 & 0.55 &      &       \\ 
020124  & \nod  & STIS  & CL    & 7418  & 2002 Apr 6  & \multicolumn{5}{c}{not detected} \\
020127  & \nod  & STIS  & CL    & 4868  & 2002 Apr 6  & 1    & 0.29 & \nod & 24.8 & OC \\
020305  & \nod  & STIS  & CL    & 6586  & 2003 Jan 20 & 1.4  & 0.15 & \nod & 25.7 & MI, OC, TT   \\
020321  & \nod  & STIS  & CL    & 4916  & 2002 Jun 7  &      &      &      & 26.4 & MI   \\
C1      &       &       &       &       &             & 1    & 0.11 & \nod &      &	    \\
C2      &       &       &       &       &             & 1    & 0.16 & \nod &      &      \\
C3      &       &       &       &       &             & 1    & 0.11 & \nod &      &      \\
C4      &       &       &       &       &             & 1    & 0.06 & \nod &      &      \\
020322  & \nod  & STIS  & CL    & 11375 & 2002 Jun 5  & \multicolumn{5}{c}{not detected} \\
020331  & \nod  & STIS  & CL    & 7202  & 2002 Aug 18 & 1    & 0.08 & \nod & 25.9 & BL \\
020405  & 0.698 & WFPC2 & F555W & 5900  & 2002 Aug 23 &      &      &      & 22.6 & AS, MI, TT \\
020405  &       & WFPC2 & F702W & 5900  & 2002 Aug 23 &      &      &      & 21.9 &       \\
020405  &       & WFPC2 & F814W & 3900  & 2002 Jun 9  &      &      &      & 21.6 &       \\ 
020410  & \nod  & STIS  & CL    & 8283  & 2003 Apr 18 & 1.1  & 0.23 &      & 24.3 & D?  \\
020410  &       & ACS   & F606W & 3680  & 2002 Jul 24 & 1.2  & 0.26 & \nod & 24.2 &          \\
020410  &       & ACS   & F814W & 3680  & 2002 Jul 24 & 1.4  & 0.24 & \nod & 23.5 &          \\
020427  & \nod  & STIS  & CL    & 8100  & 2002 Oct 26 & 0.7  & 0.46 & \nod & 24.5 & D, MI, TT  \\ 
020813  & 1.254 & ACS   & F435W & 2020  & 2003 Jul 21 & 3.4  & 0.10 & 0.84 & 24.4 & OC, TT    \\
        &       &       &       &       &             & 1    & 0.08 & 0.63 &      &          \\
020813  &       & ACS   & F606W & 1920  & 2003 Jul 21 & 2.2  & 0.11 & 0.92 & 24.2 &          \\
        &       &       &       &       &             & 1    & 0.10 & 0.84 &      &          \\
020813  &       & ACS   & F814W & 3980  & 2003 Jul 21 & 2.1  & 0.13 & 1.09 & 24.0 &          \\
        &       &       &       &       &             & 1    & 0.12 & 1.05 &      &          \\
020903  & 0.251 & ACS   & F606W & 1920  & 2003 Jun 30 &      &      &      & 21.0 & MI    \\
C1      &       &       &       &       &             & 1.6  & 0.11 & 0.41 &      &          \\
C2      &       &       &       &       &             & 0.7  & 0.38 & 1.48 &      &          \\
C3      &       &       &       &       &             & 1    & 0.07 & 0.27 &      &          \\
C4      &       &       &       &       &             & 0.3  & 0.11 & 0.41 &      &          \\
021004  & 2.323 & ACS   & F435W & 2040  & 2003 Jul 26 & 3.8  & 0.06 & 0.46 & 24.3 & BL, OC    \\
        &       &       &       &       &             & 1    & 0.05 & 0.42 &      &          \\
021004  &       & ACS   & F606W & 1920  & 2003 May 31 & 3.5  & 0.05 & 0.42 & 24.3 &          \\
        &       &       &       &       &             & 1    & 0.05 & 0.42 &      &          \\
021004  &       & ACS   & F814W & 1920  & 2003 Jul 26 & 6.9  & 0.06 & 0.50 & 24.4 &          \\
        &       &       &       &       &             & 1    & 0.04 & 0.33 &      &          \\
021211  & 1.006 & ACS   & F435W & 1920  & 2004 Jan 2  & 1.1  & 0.08 & 0.64 & 26.3 & BL       \\
021211  &       & ACS   & F606W & 1260  & 2004 Jan 5  & 1.2  & 0.07 & 0.56 & 25.5 &          \\
021211  &       & ACS   & F814W & 4000  & 2004 Jan 6  & 1.2  & 0.10 & 0.80 & 24.6 &          \\
030115  & \nod  & ACS   & F435W & 8800  & 2003 Jun 16 & 3.2  & 1.02 & \nod & 25.4 & OC, TT, MI  \\
        &       &       &       &       &             & 1    & 0.49 & \nod &      &          \\
030115  &       & ACS   & F606W & 2000  & 2003 Feb 10 & 1.4  & 0.54 & \nod & 24.9 &          \\
030115  &       & ACS   & F814W & 1920  & 2003 May 22 & 1.9  & 0.56 & \nod & 24.3 &          \\
030323  & 3.372 & ACS   & F606W & 3486  & 2003 Dec 29 & 1    & 0.08 & 0.58 & 27.7 & BL       \\
030329  & 0.168 & ACS   & F435W & 1920  & 2004 May 24 & 1.4  & 0.19 & 0.54 & 24.1 & OC \\
030329  &       & ACS   & F606W & 4000  & 2004 May 25 & 1.8  & 0.22 & 0.62 & 22.8 &          \\
030329  &       & ACS   & F814W & 2040  & 2004 May 24 & 2.1  & 0.28 & 0.78 & 22.7 &          \\
040924  & 0.859 & ACS   & F775W & 3932  & 2005 Feb 18 & 4.0  & 0.45 & 3.46 & 24.0 & OC       \\
        &       &       &       &       &             & 1    & 0.26 & 1.96 &      &          \\
040924  &       & ACS   & F850W & 3932  & 2005 Feb 19 & 1.1  & 0.31 & 2.39 & 23.8 &          \\
041006  & 0.712 & ACS   & F775W & 4224  & 2005 Feb 10 & 3.0  & 1.12 & 8.09 & 24.2 & AS       \\
        &       &       &       &       &             & 1    & 0.42 & 3.06 &      &          \\
041006  &       & ACS   & F850W & 4224  & 2005 Feb 11 & 1.1  & 0.65 & 4.68 & 24.1 &          \\
\enddata
\tablecomments{\hst\ data and morphological information for GRB host 
galaxies.  The columns are (left to right): (i) GRB name, (ii)
redshift, (iii) instrument, (iv) filter, (v) total exposure time of
drizzled image, (vi) date of observation, (vii) Sersic $n$ parameter
from GALFIT; note that a value of 1 indicates that $n$ was fixed as an
exponential profile, while multiple entries indicate a fit with $n$ as
a free and fixed parameter, (viii) effective radius in arcsec from
GALFIT, (ix) effective radius in kpc from GALFIT (using $h=0.7$), (x)
AB magnitude, (xi) morphological classification (see \S\ref{sec:morph}
for definitions).}
\end{deluxetable}

\clearpage
\begin{deluxetable}{rccccccccc}
\tabletypesize{\scriptsize}
\tablecolumns{10}
\tabcolsep0.08in\footnotesize
\tablewidth{0pc}
\tablecaption{Frequency of Morphological Properties in GRB Host Galaxies 
\label{tab:morph}}
\tablehead {
\colhead {}       		&
\colhead {AS}		        &
\colhead {KN} 		&
\colhead {OC}  		&
\colhead {TT}     	&
\colhead {MI}  		&
\colhead {DI}   		&
\colhead {BL} 		&
\colhead {TF}	 	&
\colhead{total galaxies}
}
\startdata
$z<1.0$ & 2 & 2 & 3 & 1 & 4 & 5 & 3 & 0 & 14 \\
$z>1.0$ & 4 & 3 & 2 & 4 & 5 & 1 & 5 & 2 & 16 \\
unknown $z$ & 2 & 1 & 4 & 4 & 4 & 3 & 1 & 2 & 12 \\
Total & 8 & 6 & 9 & 9 & 13 & 9 & 9 & 4 & 42 \\
\enddata
\tablecomments{Summary of morphological classification for the GRB
host galaxy sample (see \S\ref{sec:morph} for definitions).}
\end{deluxetable}

\clearpage

\clearpage

\clearpage

\clearpage

\clearpage

\clearpage

\clearpage

\end{document}